\documentclass[12pt]{article}
\usepackage{graphicx}
\begin{document}

\centerline {\large \bf Binary and Multivariate Stochastic Models
of Consensus Formation}

\bigskip
\centerline{Maxi San Miguel, V\'{\i}ctor M. Egu\'{\i}luz, and
Ra\'ul Toral}

\centerline{IMEDEA (CSIC-UIB), Campus Universitat Illes Balears,
E-07122 Palma de Mallorca, Spain}

\bigskip

\centerline{Konstantin Klemm}

\centerline{Department of Bioinformatics, University Leipzig,
Hartelstr. 16-18, 04107 Leipzig, Germany}

\bigskip

\centerline {{\bf Introduction}}

A current paradigm in computer simulation studies of social
sciences problems by physicists is the emergence of consensus
\cite{sznajd,Deffuant,Galam02,Stauffer03,Stauffer04b,Tessone04}.
The question is to establish when the dynamics of a set of
interacting agents that can choose among several options
(political vote, opinion, cultural features, etc.) leads to a
consensus in one of these options, or when a state with several
coexisting social options prevail. The latter is called a
polarized state. An important issue is to identify mechanisms
producing a polarized state in spite of general convergent
dynamics. When the agents are spatially distributed this problem
shares many characteristics with the problem of domain growth in
the kinetics of phase transitions \cite{Gunton83}: Consensus
emerges when a single spatial domain grows occupying the whole
system, while polarization corresponds to a situation in which the
system does not order and different spatial domains compete.

We consider here stochastic dynamic models naturally studied by
computer simulations. We will first review some basic results for
the voter model \cite{Liggett85}. This is a binary option
stochastic model, and probably the simplest model of collective
behavior. We focus on the dynamical effect of {\it who interacts
with whom}, that is on the consequences of different networks of
interaction. The fact whether consensus is reached or not depends
on characteristics of the network such as dimensionality. Next we
consider a model proposed by Axelrod \cite{Axelrod97} for the
dissemination of culture. This model can be considered as a
multivariable elaboration of the voter model dynamics. Time scales
of evolution scale with system size in this model in the same way
as for the voter model. We also discuss for this model the role of
different networks of interaction. Finally we consider the effect
of exogenous stochastic perturbations that account for cultural
drift.

\bigskip

\centerline{{\bf Voter model}}

The voter model
\cite{Liggett85,Frachebourg,Krapivsky92,Dornic01,Castellano03,Suchecki04,Redner04,Suchecki05}
is defined by a set of ``voters" with two opinions or spins $s_i=
\pm 1$ located at the nodes of a network. The elementary dynamical
step consists in randomly choosing one node (asynchronous update)
and assigning to it the opinion, or spin value, of one of its
nearest neighbors, also chosen at random. This mechanism of
opinion formation reflects complete a lack of self-confidence of
the agents. It could be appropriate for describing processes of
opinion formation in certain groups of teenagers in which
imitation is prevalent. The dynamical rule implemented here
corresponds to a \emph{node-update}. An alternative dynamics is
given by a \emph{link-update} rule in which the elementary
dynamical step consists in randomly choosing a pair of nearest
neighbor spins, {\it i.e.} a link, and randomly assigning to both
nearest neighbor spins the same value if they have different
values, and leaving them unchanged otherwise. These two updating
rules are equivalent in a regular lattice, but they are different
in a complex network in which different nodes have different
number of nearest neighbors \cite{Suchecki04}.

The voter model dynamics has two absorbing states, corresponding
to situations in which all the spins have converged to the $s_i=
1$ or to the $s_i= - 1$ consensus states. The ordering dynamics is
stochastic and driven by interfacial noise. This is different of
the ordering dynamics of a Glauber kinetic Ising model which is
driven by minimization of surface tension. A standard order
parameter to describe the ordering process
\cite{Dornic01,Castellano03} is the average of the interface
density $\rho$, defined as the density of links connecting sites
with different spin value. In a disordered configuration with
randomly distributed spins $\rho \simeq 1/2$, while for a
completely ordered system $\rho  \simeq 0$. In regular lattices of
dimensionality $d \leq 2$ the system orders. This means that, in
the limit of large systems, there is a coarsening process with
unbounded growth of spatial domains of one of the absorbing
states: consensus is reached. The asymptotic regime of approach to
the ordered state is characterized in $d=1$ by a power law
$\langle \rho \rangle \sim t^{-\frac{1}{2}}$, while for the
critical dimension $d=2$ a logarithmic decay is found $\langle
\rho \rangle \sim (\ln t)^{-1}$ \cite{Frachebourg,Dornic01}. Here
the average $\langle \cdot \rangle$ is an ensemble average.

\begin{figure}[hbt]
\begin{center}
\includegraphics[angle=-0,scale=0.7]{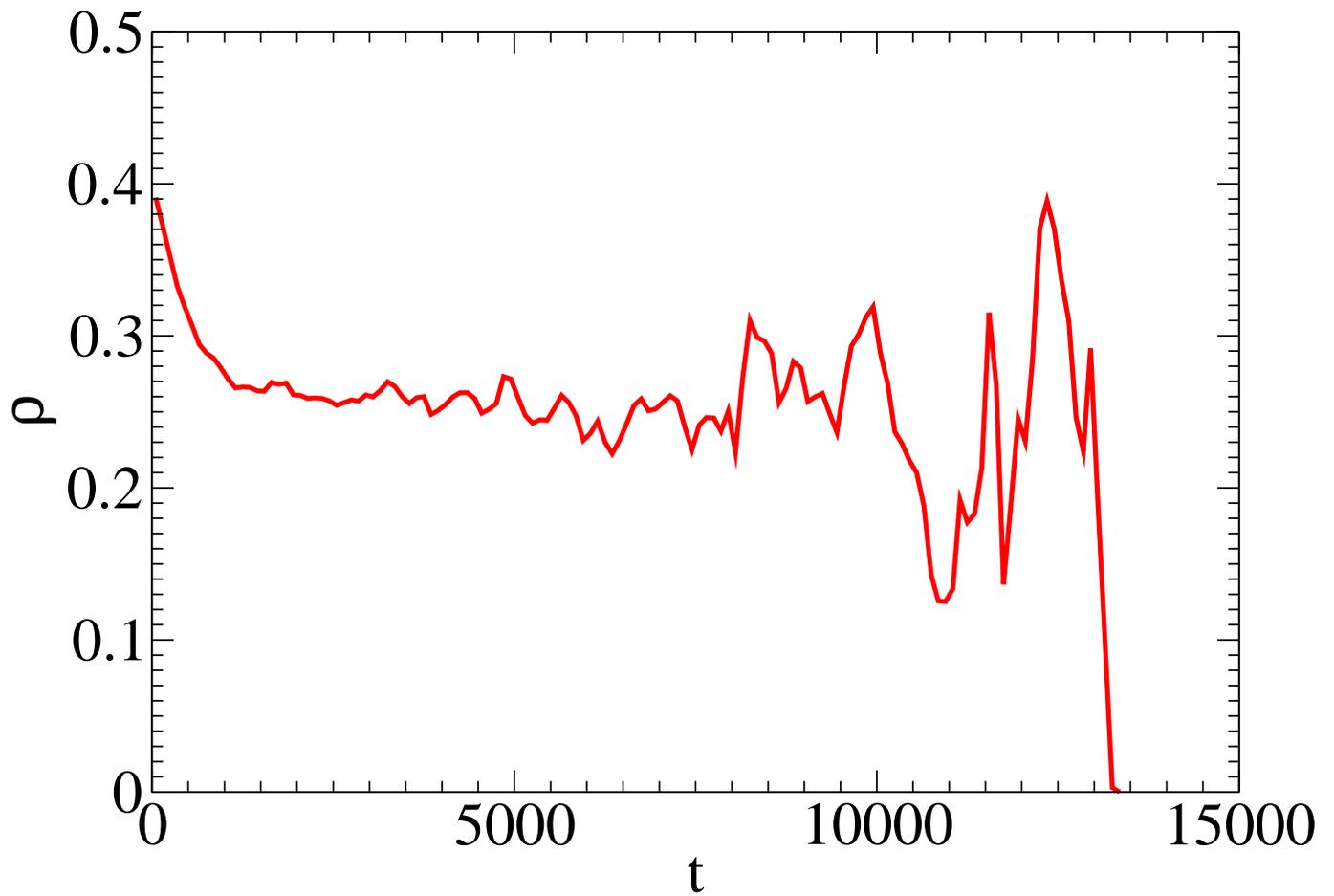}
\end{center}
\caption{Interface density evolution for an individual realization
in a scale-free Albert-Barabasi network with $N=10000$ nodes and
average connectivity $<k>=8$.}
\end{figure}

In regular lattices with $d>2$ , as well as in small-world
networks \cite{Watts98}  and scale-free Barab\'asi-Albert networks
\cite{Barabasi}, the voter dynamics does not order the system in
the thermodynamic limit of large
systems\cite{Krapivsky92,Castellano03,Suchecki04}. Starting from a
random initial condition and after an initial transient, the
system falls in a metastable partially ordered state. In the
initial transient of a given realization of the process, $\rho$
initially decreases, indicating a partial ordering of the system.
After this initial transient $\rho$ fluctuates randomly around an
average plateau value. In a finite system the metastable state has
a finite lifetime: a finite size fluctuation takes the system from
the metastable state to one of the two ordered absorbing states.
In this process the fluctuation orders the system and $\rho$
changes from its metastable plateau value to $\rho=0$ (see Fig.
1). The lifetime $\tau$ of the metastable state, for a regular
lattice in $d=3$ \cite{Krapivsky92} and also for a small-world
network \cite{Castellano03}, scales linearly with the system size
$N$, $\tau \sim N$, while a scaling $\tau \sim N^{0.88}$ has been
found \cite{Suchecki04} for the voter model in the scale-free
Barab\'asi-Albert network. The fact that a large system does not
order in a small-world or scale-free network could seem
counter-intuitive: one might argue that long distance links
(small-world) or nodes with a large number of links (hubs in a
scale-free network) should be instrumental in ordering the system.
A counter argument is that what is observed corresponds to a
network of large dimensionality: these complex networks have an
effective infinite dimension since the average path length between
two nodes grows logarithmically (or slower) with the system size.

In order to understand the different role of dimensionality and
degree distribution, i.e., the probability for a node having $k$
links (degree), one can consider the voter dynamics in the
Structured scale-free (SSF) network introduced in
Ref.~\cite{victornet1}. The SSF networks are scale-free, with a
degree distribution $P(k)\sim k^{-3}$ but are effectively one
dimensional since the average path length scales linearly with
system size $L \sim N$. Results of simulations shown in Fig. 2
indicate that the dynamics of the voter model in the SSF network
or in a regular $d=1$ network is essentially the same: the system
orders with the average interface density decreasing with a power
law with characteristic exponent $1/2$. This identifies
dimensionality and not degree distribution as the relevant
parameter to classify different classes of ordering dynamics of
the voter model in complex networks.

\begin{figure}[hbt]
\begin{center}
\includegraphics[angle=-0,scale=0.7]{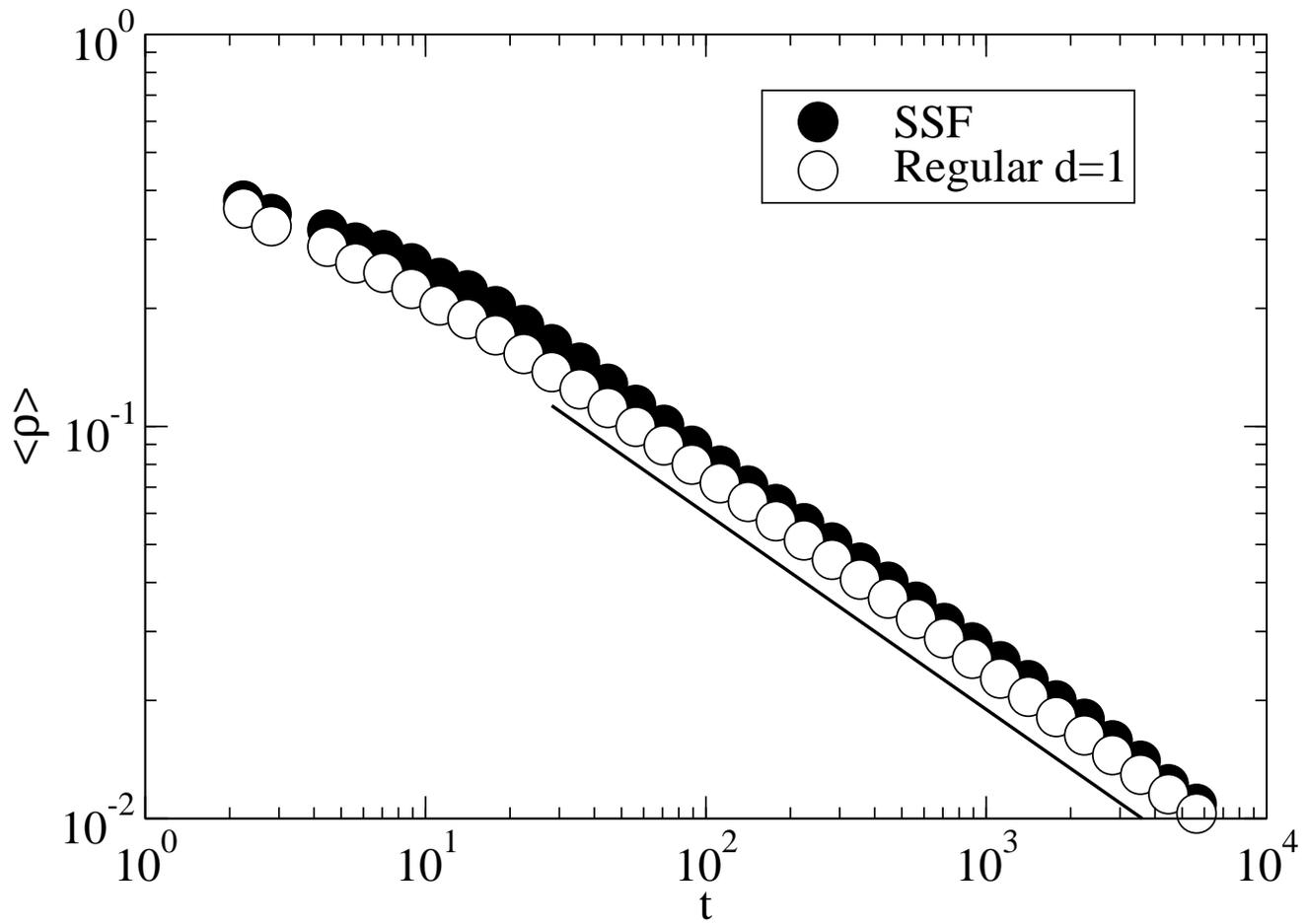}
\end{center}
\caption{Mean interface density evolution in a regular $d=1$
network and in a Structured scale-free network as indicated. The
average is over 1000 realizations. $N=10000$ and $<k>=8$. The
continuous line indicates a power law decay with exponent $-1/2$.}
\end{figure}

The voter model can  also be studied in other different complex
networks of dimension $d>1$ characterized by a parameter $p$
measuring the disorder of the network. This parameter is the one
originally used to characterize a small-world network
\cite{Watts98}, varying continuously from $p=0$ (regular network)
to $p=1$ (random network). One finds that network disorder
decreases the lifetime of the metastable disordered states.
Likewise, the lifetime of these states is decreased when the
networks have nodes with a large number of links
\cite{Suchecki05}.

\bigskip

\centerline{{\bf Axelrod model}}

Axelrod \cite{Axelrod97} addressed the issue of the persistence of
cultural diversity asking the following question: {\it if people
tend to become more alike in their  beliefs, attitudes and
behavior when they interact, why do not all differences eventually
disappear?"} To answer this question he proposed a model to
explore mechanisms of competition between globalization
(consensus) and coexistence of several cultural options
(polarization). The basic premise of the model is that the more
similar an actor is to a neighbor, the more likely the actor will
adopt one of the neighbor's traits. In addition to treating
culture as multidimensional (not a binary option), a novelty of
the model is that its dynamics takes into account the interaction
between the different cultural features. The model is defined by
considering $N$ agents as the nodes of a network of interaction.
The state of agent $i$ is a vector of $F$ components ({\em
cultural features})
$(\sigma_{i1},\sigma_{i2},\cdots,\sigma_{iF})$. Each $\sigma_{if}$
is one of the $q$ integer values ({\em cultural traits})
$1,\dots,q$, initially assigned independently and with equal
probability $1/q$. The time-discrete dynamics is defined as
iterating the following steps:
\begin{enumerate}
\item Select at random a pair of sites of the network connected by a
link $(i,j)$.
\item Calculate the {\em overlap} (number of shared features $\sigma_{ik}=\sigma_{jk}$.)
$l_{ij}$.
\item If $0<l_{ij}<F$, the link is said to be {\sl active} and sites
$i$ and $j$ interact with probability $l_{ij}/F$ (similarity
rule). In case of interaction, choose $g$ randomly such that
$\sigma_{ig}\neq\sigma_{jg}$ and set $\sigma_{ig}=\sigma_{jg}$.
\end{enumerate}

The model has $q^F$ equivalent cultural options. Consensus (global
culture) is reached if a domain of one of these options occupies
the whole system. For $q=2$ Axelrod's model can be viewed as a set
of $F$ coupled voter models. For a general value of $q$ it still
shares with the voter model the basic stochastic dynamics driven
by interfacial noise as shown in Fig. 3: An initial condition of a
bubble of one of the $q^F$ cultures on the background of another
cultural option with only one feature in common dissolves by
interfacial noise. Several snapshots of the dynamical evolution
from random initial conditions in a $d=2$ square lattice are shown
in Fig. 4.\cite{java} For a given value of $F$ the evolution from
initial random conditions leads to a state of global culture
(consensus) or a multicultural state depending on the value of
$q$. The parameter $q$ is a measure of the degree of initial
disorder in the system. The fact that multicultural disordered
states are reached illustrates how local convergence, enforced by
the similarity rule used in the dynamics, can generate global
polarization.

\begin{figure}[hbt]
\begin{center}
\includegraphics[angle=-0,scale=0.22]{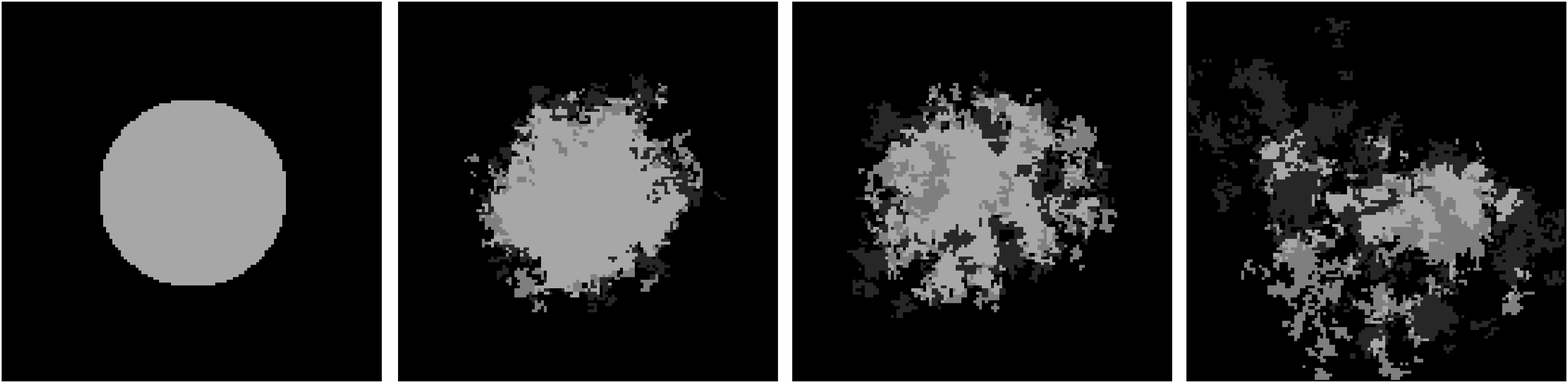}
\end{center}
\caption{Snapshots of the time evolution of Axelrod model at times
$t=0, 114, 272, 1331$. Different colors indicate different
cultural states. System size $N=128 \times 128$. Parameter values
$F=3$, $q=15$.}
\end{figure}

\begin{figure}[hbt]
\begin{center}
\includegraphics[angle=-0,scale=0.22]{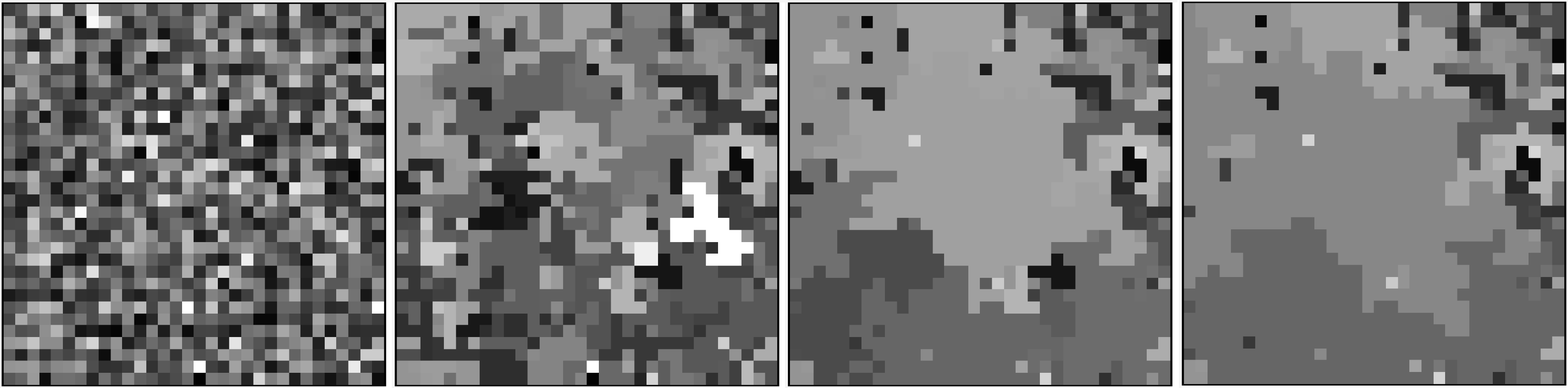}
\end{center}
\caption{Snapshots of the time evolution of Axelrod model from
random initial conditions at times $t=0, 1000, 3000, 6807$. At
time $t=6807$ the dynamics stops and the configuration is frozen.
System size $N=32 \times 32$. Parameter values $F=3$, $q=10$. }
\end{figure}

A systematic analysis of the dependence on $q$ can be carried out
from the point of view of Statistical Physics \cite{Castellano00}
through Monte Carlo computer simulations. Defining an order
parameter as the mean value of the relative size of the largest
homogeneous cultural domain $S_{\rm max}$, one finds a
nonequilibrium order-disorder transition as shown in Fig. 5 for a
$d=2$ square lattice: There exists a threshold value $q_c$, such
that for $q<q_c$ the system orders in a consensus monocultural
uniform state ($<S_{\rm max}>/N \sim 1)$, while for $q>q_c$ the
system freezes in a polarized or multicultural state ($<S_{\rm
max}>\ll N$). The transition becomes sharp and well defined for
large systems and it is a first-order transition in $d=2$, while
it becomes a continuous transition in $d=1$
\cite{Klemm03c,Vilone02,Klemm05}. In $d=1$ the Axelrod dynamics is
an optimization dynamics for which a Lyapunov potential can be
found \cite{Klemm05}. We note that $F=2$ is a special case
\cite{Castellano00,Vilone02} that we do not discuss here. We also
note that $q_c$ and the transition itself is defined considering
the dynamical evolution form an initial random disordered
configuration and not for arbitrary initial conditions. We use
here a set of uniform random initial conditions, while other
authors have used a a Poisson distribution for the initial random
values of $q$ \cite{Castellano00}.

\begin{figure}[hbt]
\begin{center}
\includegraphics[angle=-0,scale=1]{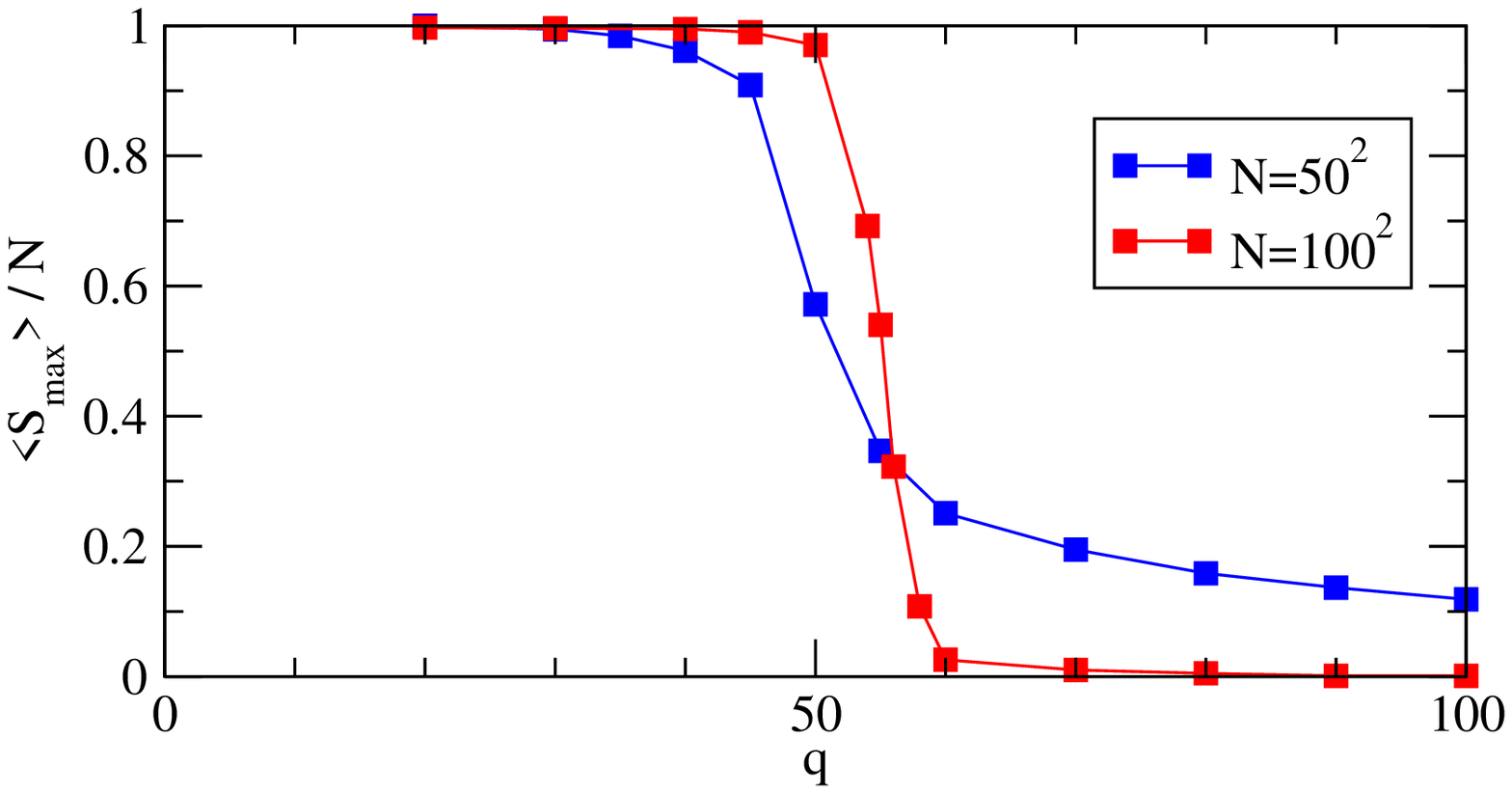}
\end{center}
\caption{Normalized order parameter $<S_{\rm max}>/N$ as a
function of $q$ for $d=2$ square lattices of sizes $N=50 \times
50$ and $N=100 \times 100$ for $F=10$.}
\end{figure}

\bigskip

\centerline{{\bf Axelrod model in complex networks}}

The network of interactions among the agents accounts for the
local geography in Axelrod's model. Following our discussion of
the voter model, it is natural to ask how the above results for a
regular network are modified when considering a complex network of
interaction \cite{Klemm03b}. An expectation is that with random
long distance interactions, the heterogeneity sustained by local
interactions can no longer be maintained \cite{Axelrod97}. For a
small-world network it is found that the transition remains
sharply defined as the system size increases, but it is shifted to
larger values of $q$ as the disorder parameter $p$ is increased.
So that, as expected, small-world connectivity favors cultural
globalization. This is shown in the phase diagram of Fig. 6 in
which we observe that for a given value of $q$ in which the system
is in a polarized state in regular network, consensus (global
culture) can be reached by increasing the disorder parameter of
the network, $p$.

\begin{figure}[hbt]
\begin{center}
\includegraphics[angle=-0,scale=1]{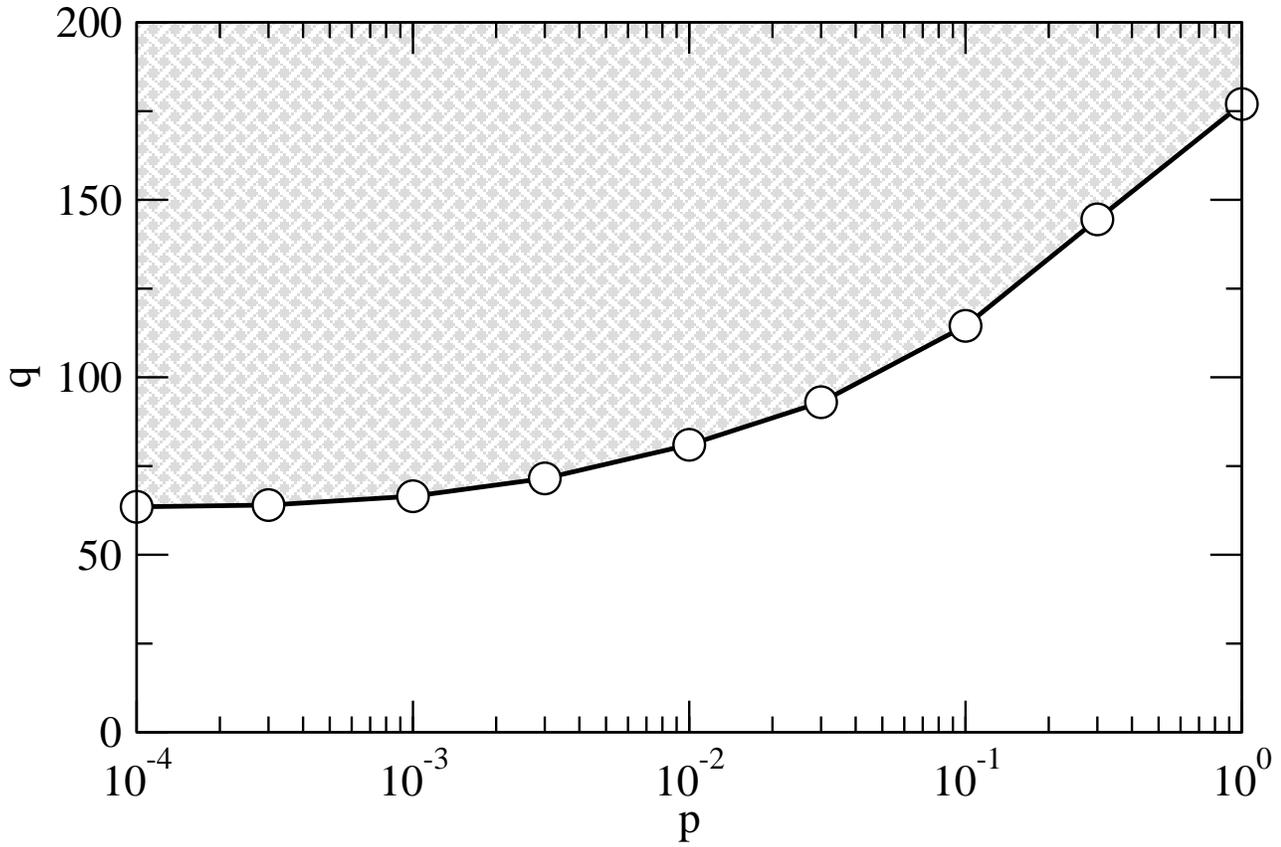}
\end{center}
\caption{Phase diagram for the Axelrod model in a small-world
network of size $N=500^2$ for $F=10$. The shaded area are $(q,p)$
parameters for which a polarized or multicultural state is
reached. The other side of the continuous curve corresponds to
parameters for which consensus (state of global culture) is
reached \cite{Klemm03b}.}
\end{figure}

In a scale-free Barab\'asi-Albert network \cite{Barabasi} the
order-disorder transition of the Axelrod model becomes system-size
dependent and the critical value $q_c$ is shifted to larger and
larger values as $N$ increases, so that a state of global culture
(consensus) prevails in the limit of large systems. In addition,
for a fixed large value of $N$ and fixed average connectivity
$<k>$, $q_c$ is larger in a scale-free network than the limiting
value of $q_c$ found for $p=1$ in a small-world network: The scale
free connectivity is more efficient than a random connectivity
($p=1$) in promoting global culture. These results for the Axelrod
model in small-world and scale-free networks parallel what happens
for a kinetic Ising model: the small-world connectivity increases
the critical temperature, while the critical temperature diverges
with system size in a scale-free network.

Similarly to the discussion of the voter model, we can ask here
about the specific role of the degree distribution in the fact
that the transition disappears for a  large systems in a
scale-free Barab\'asi-Albert network. Considering again the
Structured scale-free (SSF) network introduced in
Ref.~\cite{victornet1} we find that the transition remains here
well defined at a finite value of $q$ for large systems. The
conclusion is that it is the spatial dimensionality of the
interaction network, and not just the presence of hubs, what gives
rise to the divergence of $q_c$ with $N$. On the other hand, hubs
create local order in the system so that for the multicultural
disordered state in a SSF network $<S_{\rm max}>$ takes a finite
value.

\bigskip

\centerline{{\bf Cultural drift: Exogenous perturbations in
Axelrod model}}

Among the open questions discussed by Axelrod in his original work
\cite{Axelrod97} he mentions that  {\it Perhaps the most
interesting extension and at the same time, the most difficult one
to analyze is cultural drift}, and he suggests to model it as
spontaneous changes of cultural traits. Cultural drift takes into
account that there is always some influence between neighbors even
when they have completely different cultures. In the language of
physics simulations he is asking about the role of noise in the
order-disorder transition discussed above. The stochastic dynamics
giving rise to this transition can be considered as a zero
temperature dynamics. The question is if this transition is robust
against the presence of fluctuations, or if any finite fluctuation
disorders the system, as it happens in the $d=1$ kinetic Ising
model. Generally speaking noise is known to have two different
effects, one is to produce disorder by accumulation of
fluctuations, but another one is to help the system in finding
paths in which it can escape from frozen disordered
configurations, leading to ordered states. An alternative way of
formulating the question is then if external perturbations acting
on a frozen  multicultural state can take the system to the
consensus state. To address these issues we implement cultural
drift in the model adding a fourth step in the iterated loop of
the dynamics defined above \cite{Klemm03a}:

4. With probability $r$, perform a single feature
perturbation in which randomly choosing an agent $i$ and one of
its features $f$, the trait $\sigma_{if}$ is replaced by a new
randomly chosen value.

\begin{figure}[hbt]
\begin{center}
\includegraphics[angle=-0,scale=1]{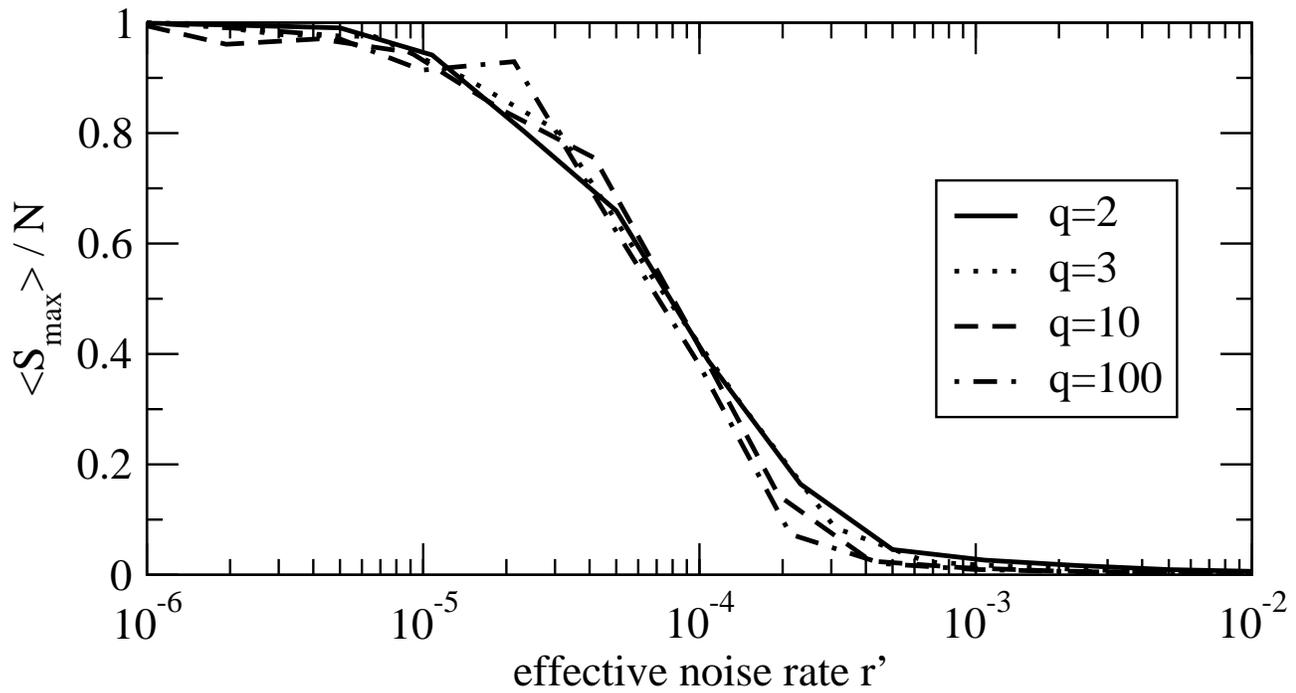}
\end{center}
\caption{Normalized order parameter $<S_{\rm max}>/N$ as a
function of the effective noise rate $r^{'}$ for different values
of $q$ in a $d=2$ square lattice of size $N=50 \times 50$ and
$F=2$ \cite{Klemm03a}.}
\end{figure}

Simulation results for a $d=2$ square lattice are shown in Fig. 7:
We observe a transition from multicultural to consensus states
controlled by an effective noise rate $r^{'}=r(1-1/q)$. The factor
$(1-1/q)$ takes into account the probability that the single
feature perturbation does not change the value of the trait. This
is a noise induced transition since the control parameter is a
noise property. In addition, the transition has universal scaling
properties with respect to $q$: the same result is found for
different values of $q$ and a consensus state is found for any
value of $q$ as $r$ goes to zero. Therefore, cultural drift
destroys the transition controlled by $q$ that was found in the
absence of exogenous perturbations ($r=0$). In this sense, noise
is here an essential parameter that changes completely the type of
transition exhibited by the system. An additional important point
is the character of the states found at both sides of the noise
induced transition. The disordered multicultural state found for
large $r$ is no longer a frozen configuration, but rather it
exhibits disordered noise-sustained dynamics. On the other hand,
the consensus or ordered state found for small $r$ is metastable:
Once one of the equivalent $q^F$ cultural states is reached, the
systems does not stay there forever, but eventually a fluctuation
takes it from this state to another one of the equivalent  $q^F$
states, as shown in Fig. 8.

\begin{figure}[hbt]
\begin{center}
\includegraphics[angle=-0,scale=0.22]{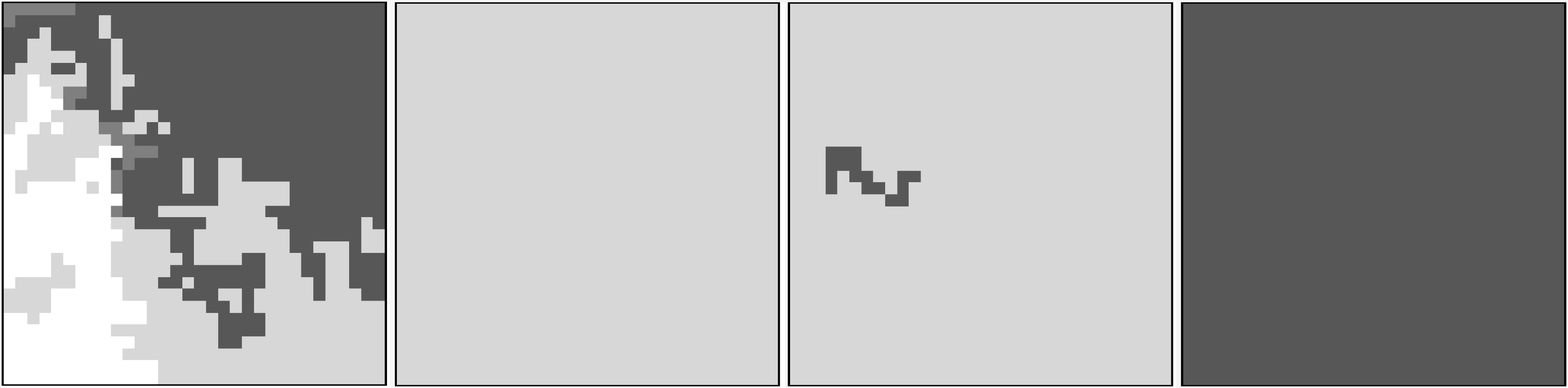}
\end{center}
\caption{Snapshots of the time evolution of Axelrod model with
exogenous perturbations in a $d=2$ square lattice of size $N=32
\times 32$ with $F=3$, $q=2$ and $r=0.000017$. A random
configuration is chosen at the initial time $t=0$. Snapshots are
shown at times $t=1650, 5519, 180000, 204000$. At time $t=1650$
the system is evolving to a metastable consensus state reached at
$t=5519$. The system remains there for a long time until a large
enough fluctuation of another equivalent consensus state occurs
($t=180000$) and takes the system to that state ($t=204000$). }
\end{figure}

Why does the noise rate cause a transition? There is here a
competition between two time scales, the time scale at which noise
is acting ($1/r$) and the relaxation time of perturbations $T$.
For small noise rate $r$ there is time to relax and the system
decays to a consensus state, while for a large noise rate,
stochastic perturbations accumulate and multicultural disorder is
built up. The transition is then expected to occur for $rT \sim
1$. The relaxation time $T$ of perturbations can be calculated as
an exit time in a random walk \cite{Klemm05,Klemm03a}. In a mean
field approximation it is given as the time needed to reach
consensus in a finite system following the voter model dynamics.
For a $d=2$ square lattice this is $T \sim N \ln N$
\cite{Krapivsky92,Klemm03a}. The noise induced transition occurs
then for a system size dependent value of $r$, but curves as the
ones plotted in Fig. 8 for different values of $N$ collapse into a
single curve when plotted versus $rN \ln N$ \cite{Klemm03a}. The
general result is that in the limit of very large systems,
disordered multicultural states prevail at any noise rate.
Therefore cultural drift causes global polarization in large
systems, but as a state with noise-sustained dynamics rather than
a frozen configuration of spatially coexisting equivalent
cultures.

\bigskip

\centerline{{\bf Summary}}

We have reviewed some aspects of stochastic dynamical models of
consensus formation. The simple voter model has been used to
illustrate how this stochastic dynamics is very much affected by
the spatial background in which it takes place: Different
characteristics of the network of interactions determine if
consensus grows in the system or if a polarized disordered state
prevails. We have also considered these questions in a related
model due to Axelrod which goes beyond the usual binary options of
spin models and that also incorporates interaction among
multivalued options. For this model we have also shown that
exogenous stochastic perturbations are essential, since they
completely change the nature of the states reached by the system
in its dynamical evolution. An interesting open question for
future developments is to go beyond the static networks of
interaction considered here, allowing for a co-evolution of the
network and the state of the agents in the network. This general
idea of co-evolution has been implemented already in other
computer simulations of social dynamics \cite{prisoner2}.

\bigskip
We acknowledge the collaboration of K. Suchecki in the original
studies of the voter model dynamics. We also acknowledge financial
support from MEC (Spain) through project CONOCE2
(FIS2004-00953)

\end{document}